\documentclass{epl}
\usepackage{graphicx}

\def\a{\alpha}
\def\b{\beta}

\def\L{\Lambda}
\def\r{\rho}

\def\ra{\rightarrow}
\def\Ra{\Rightarrow}

\def\bm#1{\mbox{\boldmath{$#1$}}}

\title{The fractal distribution of haloes}

\author{Jos\'e Gaite}
\institute{Instituto de Matem{\'a}ticas y F{\'\i}sica Fundamental,
CSIC, Serrano 113bis, 28006 Madrid, Spain}


\begin{document}

\maketitle

\label{firstpage}

\begin{abstract}
We examine the proposal that a model of the large-scale matter
distribution consisting of randomly placed haloes with power-law
profile, as opposed to a fractal model, can account for the observed
power-law galaxy-galaxy correlations.  We conclude that such model,
which can actually be considered as a {\em degenerate multifractal}
model, is not realistic but suggests a new picture of multifractal
models, namely, as sets of fractal distributions of haloes.  We
analyse, according to this picture, the properties of the matter
distribution produced in cosmological $N$-body simulations, with
affirmative results; namely, haloes of similar mass have a fractal
distribution with a given dimension, which grows as the mass
diminishes. 
\end{abstract}

\section{Introduction}

The model of the large-scale matter distribution consisting of haloes
with power-law profile {\em and} with centers distributed randomly can
be traced back to Peebles \cite{Pee}. McClelland \& Silk \cite{McC-Silk} 
and later Sheth \& Jain \cite{Sheth-J} and
Murante et al \cite{Mur} studied the effect on correlation functions of a
spectrum of halo masses. The latter authors claimed that this model
provides a good description of the observed galaxy correlations, at
least on small scales.  In particular, their analysis of generalized
correlation integrals shows non-trivial scaling, which they consider
indistinguishable from the scaling in fractal models.  Recently, this
conclusion has led Botaccio, Montuori and Pietronero \cite{BMP} to
construct new statistical functions to distinguish both types of
models.

Motivated by the mentioned preceding works, here we critically examine
the model of randomly distributed haloes with power-law profile.  We
instead propose a model consisting of a set of fractal distributions
of haloes, that is, combining both the fractal and the halo models.
Actually, Murante et al already considered to place the haloes ``more
deliberately, for instance in the sheet-like structures produced by
large-scale pancake formation'' but deferred it to future work. Here,
we constructively proceed from the case of a single halo, understood
as a {\em degenerate} multifractal model, to a general multifractal,
considered as a set of fractal distributions of haloes. We argue that
this model is plausible in regard to models of structure formation and
provides a better description, especially, of {\em void} regions.
Furthermore, we show that it is supported by the multifractal analysis
of $N$-body simulations.  We also discuss the r\^ole of the two-point
correlation function in this context.

The halo model of large-scale structure, to be discussed further
below, is now well developed \cite{CoorShet}. Actually, clustering in
the distribution of haloes has been studied by Mo \& White \cite{Mo-W}
and Sheth \& Tormen \cite{Sheth-Tor}.  However, there is no study
within the multifractal formalism yet (the paper by Murante et al
\cite{Mur} relates haloes with multifractals but is restricted to
randomly distributed haloes).  The advantage of this study is that it
unveils the scaling properties and, in particular, allows us to 
deeply relate halo and fractal models.

\section{Singularities and Multifractals}

The standard magnitude used to describe fractals is the number-radius
function $N(r)$, defined as the average number of points in 
a ball of radius $r$ centred on a point:
%(integral two-point correlation function): 
it must be
a power law, namely, $N(r) \sim r^D$, where $D$ is the fractal
dimension.  As a generalization of fractal distributions, multifractal
measures represent mass distributions spread according to highly
irregular patterns, so that a local density fails to exist. In fact, a
multifractal measure is locally such that, in a ball of radius $r$
centred on a point ${\bm x}$, $\mu[B({\bm x},r)] \sim r^\a$, where
the Lipschitz-H\"older exponent $\a$ takes on a range of values
\cite{Falc}. For $\a < 3$ the local density fails to exist, that
is, there is no $\lim_{r\ra 0}\mu[B({\bm x},r)]/r^3$ (it is
divergent). On the other hand, for a differentiable distribution
$\mu[B({\bm x},r)] \sim r^3$, so $\a = 3$ and the density is
$\r({\bm x}) = \lim_{r\ra 0}\mu[B({\bm x},r)]/(\frac{4}{3}\pi r^3)$.

Let us see how this definition applies to a differentiable
distribution given by a single power law of the radius from a given
point in space (placed at the origin), namely, $\r(r) =
A\,r^{-\beta},\: 0 < \b < 3$.  This distribution is differentiable at
every point except the origin, which is a singularity.  Note that if
${\bm x} \neq 0$ then $\mu[B({\bm x},r)] \sim r^{3}$, whereas at the
origin $\mu[B(0,r)] \sim r^{3-\b}.$ The mass function $M(r)$,
namely, the average of the mass in a ball of radius $r$ is
\begin{eqnarray}
M(r) = \int d^3x\, \r({\bm x})\, \mu[B({\bm x},r)] %\nonumber\\
=  \int d^3x \,{x}^{- \beta }\int_{B({\bf x},r)} d^3s\,
         {\left|{\bm x} +  {\bm s} \right|}^{-\beta},
\label{Mr}
\end{eqnarray}
where we have to integrate over a finite region to have finite total
mass (for example, with an $\left|{\bm x}\right|$ cutoff).  This
also renders $M(r)$ finite, for there is no divergence at ${\bm x} =
{\bm s} = 0$.  The function $M(r)$ can actually be computed in terms
of trascendental functions, but this is not necessary, because only
the behaviour of $M(r)$ in the limit of $r \ra 0$ is
needed. Therefore, in the integrand of Eq.~(\ref{Mr}) only two cases
occur: either ${\bm x} \neq 0$ or ${\bm x} = 0$ (the singularity). In
the first case, the integral over ${\bm s}$ in the $r \ra 0$ limit
yields ${x}^{- \beta }\,4\pi r^3/3 \propto r^3$. In contrast, to
calculate the integral at ${\bm x} = 0$, we must perform the limit
${\bm x} \ra 0$ before the limit $r \ra 0$. The result is
approximately $\int_0^r dx\, {x}^{2- \beta }\int_0^r ds\,
{s}^{2-\beta} \propto r^{2(3-\b)}$. Note that this quantity coincides
with $\mu[B(0,r)]^2$. We conclude that if $\b < 3/2$ the behaviour of
$M(r)$ in the limit of $r \ra 0$ is $M(r) \propto r^3$ but if $\b >
3/2$ the behaviour is instead $M(r) \propto r^{2(3-\b)}$ and is
dominated by the singularity.  An important remark is that, in the
case dominated by the singularity, the value of the density $\r({\bm
x})$ at regular points (${\bm x} \neq 0$) is irrelevant and the result
is therefore {\em independent} of its power-law form at those points.

We can consider a single power-law singularity as an example of
multifractal \cite{Halsey}.  In a generic multifractal the
measure is concentrated in a set of vanishing Lebesgue measure, so
that in almost every point there is no measure.  The distribution of
the measure is characterized by the {\em multifractal spectrum}, which
can be calculated from the moment sums $M_q(r) = \sum_{\bf x}
\mu[B({\bm x},r)]^q$ (generalized correlation integrals) 
\cite{Halsey,Falc}. $M_2(r)$ is equivalent to $M(r)$ in
Eq.~(\ref{Mr}).  The larger the value of $q$, the more important is
the contribution of the most singular points to the corresponding
moment sum, and it can happen that only a few points (or a single one)
contribute from some $q$ upwards.  The multifractal spectrum $f(\a)$
is the function that gives the fractal dimension of the set of points
with exponent $\a$, and can be calculated from the moments $M_q(r)$
($r \ra 0$).

For a distribution with {\em one} singularity with exponent $\a= 3-\b$
and regular with positive density everywhere else, the singularity
contribution is $\mu[B(0,r)]^q \sim r^{q(3-\b)}$ and the regular point
contribution $\mu[B({\bm x},r)]^q \sim r^{3(q-1)}$.  Therefore, if $q
< 3/\b$ then $\a = f(\a) = 3$ and if $q > 3/\b$ then $\a = 3-\b,\:
f(\a) = 0$. This {\em degenerate} multifractal spectrum can be derived
in the standard way through the function $\tau(q)$ such that $M_q(r)
\sim r^{\tau(q)}$.  We have that $\tau(q) = 3(q-1)$ if $q < 3/\b$ and
$\tau(q) = q(3-\b)$ if $q > 3/\b$, corresponding to $\a = 3$ and $\a =
3-\b$, respectively.  They yield the fractal dimensions of the
singular set $f(\a= 3) = \left.q\a - \tau(q)\right|_{\a= 3} = 3$ and
$f(\a= 3-\b) = \left.q\a - \tau(q)\right|_{\a= 3-\b} = 0$,
respectively, as expected.

A generic multifractal is singular in an uncountable set of points (of
vanishing Lebesgue measure). However, we can have intermediate cases,
namely, distributions differentiable everywhere except in a set with
``many" points and, furthermore, such that the measure is a power-law
near every singular point.  It is crucial what meaning we shall
attribute to the word ``many". If the set of singularities is finite,
then each one is isolated. If there is a common exponent, no novelties
arise. Otherwise, depending on each singularity's strength, as given
by its exponent, $\tau(q)$ undergoes crossovers as $q$ increases,
which can mimick a continuous variation. However, this type of
multifractal is still degenerate and, in addition, $f(\a)$ jumps from
3 to 0 at once.  All this applies to the model of randomly-placed
power-law haloes \cite{Pee,McC-Silk,Mur}.

Let us now consider the generalization to a distribution that is
almost everywhere regular but with power-law singularities in a
fractal set.  Let us assume same strength singularities (common
exponent $\a= 3-\b$, $0 < \b < 3$) and everywhere non-vanishing
density.  A careless generalization of the case with a finite
singularity set would make us believe that the multifractal spectrum
is not altered and still $\tau(q) = 3(q-1)$ if $q < 3/\b$ and $\tau(q)
= q(3-\b)$ if $q > 3/\b$ (corresponding to $\a = 3$ and $\a = 3-\b$,
respectively).  However, this is now inconsistent, for it yields the
fractal dimension of the singular set $f(\a= 3-\b) = \left.q\a -
\tau(q)\right|_{\a= 3-\b} = 0$, contrary to the assumption.  So
$\tau(q) \neq q(3-\b)$ for $q > 3/\b$.  In fact, given that $\a(q)=
\tau'(q)$, $\a$ constant only implies that $\tau(q) = \a\,q + c$, with
$c$ being some arbitrary constant. Furthermore, $\tau(q) = \a\,q + c
\Ra f(\a) = -c$, so any dimension of the fractal singularity set $D =
-c$ is allowed.  Assuming $\tau(q)$ continuous, its change of
behaviour takes place at $q = (3-D)/(3-\a)$, smaller than for a finite
number of singularities of common strength (but larger than unity).
The reason for the smaller value of $\tau(q)$ for $q > (3-D)/(3-\a)$
is that almost every point of a fractal is an accumulation point
(because isolated points do not contribute to the fractal dimension),
so the singularities are enhanced.

Summarizing, the generalization from a finite set of common strength
singularities to a dimension-$D$ fractal set of common strength
singularities leads to a crossover at smaller $q$ ($q \geq 1$) and
with smaller (more singular) $\tau(q) = \a\,q - D$. A crossover
between just two exponents $\a$ (one of which may be the trivial $\a =
3$ or not) corresponds to a {\em bifractal}.  Arguments for bifractal
distributions of matter in cosmology have been given by Balian and
Schaeffer \cite{Bal-Schaf}, and Borgani \cite{Borga} has shown that
these distributions can be derived from the generalized
thermodynamical model.

We can have several values of the exponent $\a$, like in the case of a
finite singularity set.  Therefore, we can have several crossovers,
defining a multifractal spectrum degenerate by pieces.  Let us
consider the definition $D(q) = \tau(q)/(q-1)$, and the entropy
dimension $D(1) = \a = f(\a)$, which defines the set of the measure's
concentrate. A monofractal is most degenerate, in the sense that there
is only one $\a$, such that $f(\a) = \a$, of course.  There are less
degenerate cases in which $\a(q)$ is constant by pieces.  In them the
measure's concentrate still has $f(\a) = \a$, but the other fractal
sets with constant $\a$ have $f(\a) < \a$. To probe these sets one has
to measure moments with the appropriate $q$'s.

Of course, a multifractal spectrum degenerate by pieces approaches a
non-degenerate one as the number of pieces grows and the exponent
$\a(q)$ becomes continuous. Usual physical mechanisms that generate
singularities produce this type of spectrum.  For example, the set of
singularities of a mass distribution may be determined by some random
process, such as the maxima of fractional Brownian motion. Models of
this type are relevant in the context of the halo model of large-scale
structure.

\section{The halo model of large-scale structure}

The halo model of large-scale structure takes its inspiration from old
ideas on gravitational clustering of matter and also from the analysis
of $N$-body simulations \cite{CoorShet}. It assumes that the
dark matter is in the form of collapsed and virialized haloes with
definite density profile.  This profile is usually taken to have
spherical symmetry around the halo centre (with small deviations),
with a given radial profile. This radial profile can adopt several
forms, the simplest one being a power law $\r(r) \propto r^{-\b}.$

Complementary to the profile is the distribution of halo centres. Of
course, it is simpler to consider a finite number of well-separated
randomly-distributed haloes, but the usual mechanisms for large-scale
structure formation produce an infinite number, unless a {\em lower
cutoff} is introduced. This lower cutoff may appear as a softening 
of the gravitational force or as a minimal mass (both are present 
in $N$-body simulations). 
A lower cutoff prevents infinite clustering on
small scales but nonetheless preserves clustering on larger scales.
Let us focus on two models of large-scale structure formation, namely,
the spherical collapse model and the adhesion model.

The spherical collapse model can be formulated in several ways; we
here consider the formulation that takes the peaks of the initial
Gaussian density field as seeds for the formation of haloes, so the
number of haloes is directly determined by these peaks, although the
spatial distribution and final halo masses depend on the ensuing
dynamics. This nonlinear dynamics is complex but nonetheless enhances
clustering. In any event, the set of peaks of a Gaussian density field
is {\em dense}, so the assumption of a finite number of haloes is
clearly insufficient. Arguably, this dense set of points becomes a set
of fractals under the nonlinear dynamics, but this model does not
allow to conclude much more without further assumptions on that
nonlinear dynamics.

The adhesion model \cite{adhesion}
provides us with a concrete nonlinear equation and
methods to study the evolution of an initial Gaussian density field
\cite{Verga}. Actually, the evidence suggests the
development of a distribution with multifractal
features. It consists of a self-similar distribution of caustics:
pancakes, filaments and nodes.  Unfortunately, this model only covers
the early stages of structure formation, before the virialization of
haloes, which are supposed to arise essentially from the caustic
nodes.  The resulting distribution is probably similar to what Murante
et al \cite{Mur} had in mind as singularities ``placed more deliberately,
for instance in the sheet-like structures produced by large-scale
pancake formation''.

Since the theoretical models of large-scale structure formation are
insufficient to fully determine the structure and distribution of
haloes, we may turn to $N$-body simulations. In fact, a good deal of
evidence for the halo model comes from them. However, discretization
issues tend to make the conclusions less certain.  For example, the
inevitable smoothing length that is introduced in the gravitational
interaction suppresses clustering on smaller scales, so that the
smoothness of haloes may be just a consequence of it rather than a
consequence of the gravitational dynamics. At any rate, the smoothness
of haloes is questionable and there is evidence for a hierarchical
structure of condensations, similar to a multifractal \cite{Vala}.

Multifractal analyses of large-scale structure were first introduced
regarding the distribution of galaxies \cite{multif1,multif2} and
later applied to $N$-body simulations \cite{Colom,Valda,Yepes}, with
positive results. However, the relationship of these results with
haloes and their distribution is far from clear.  One way to clarify
this relationship is to give a definition of haloes suitable for
$N$-body simulations and then analyse their spatial distribution. In a
multifractal model, haloes must be identified with density
singularities, but there are no real singularities in $N$-body
simulations.  One can define a coarse-grained density and consider
density peaks as singularities, in accord with {\em coarse 
multifractal analysis} \cite{Falc}.  
Then one can measure the strength of
singularities and analyse the distributions of singularities of
similar strength, to assess their fractality.  This is an alternative
(or rather complementary) procedure to the computation of moments and
hence the multifractal spectrum $f(\a)$.

\begin{figure}
%\centering
\begin{minipage}{7.5cm}
\includegraphics[width=7cm]{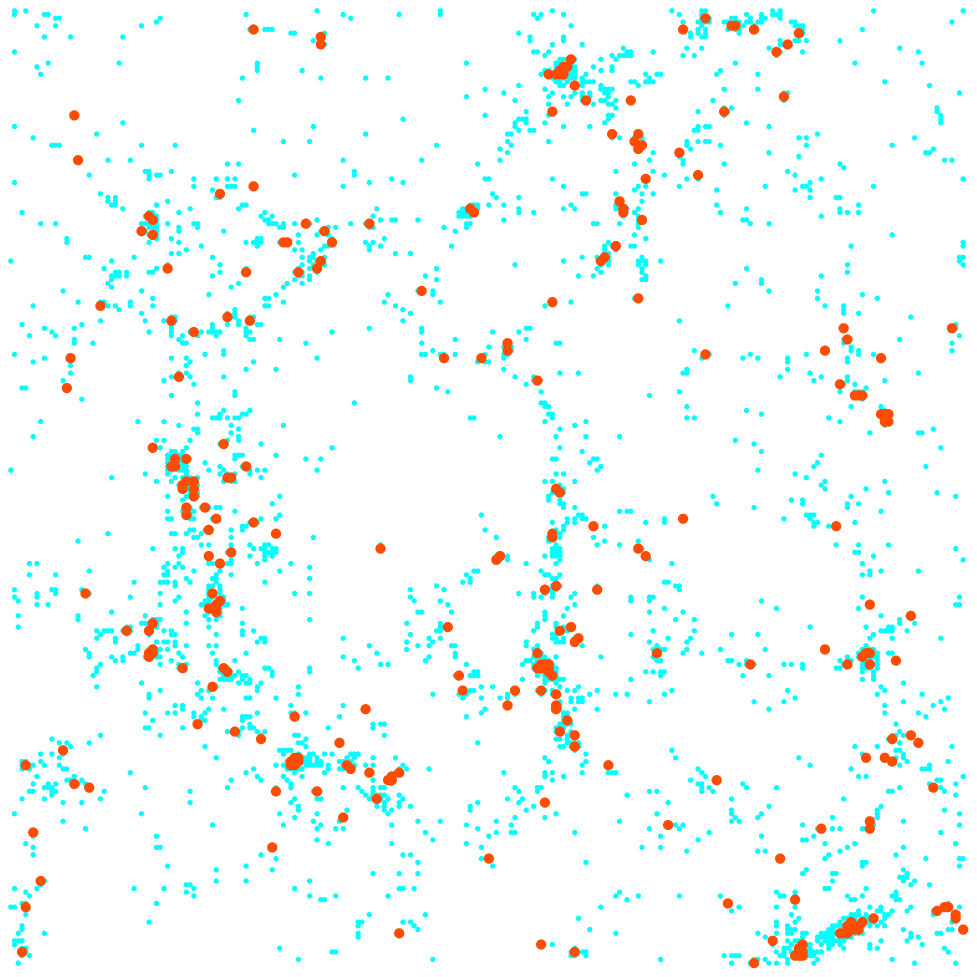}
\end{minipage}
%\vspace{2mm}
\begin{minipage}{7.5cm}
\includegraphics[width=7.5cm]{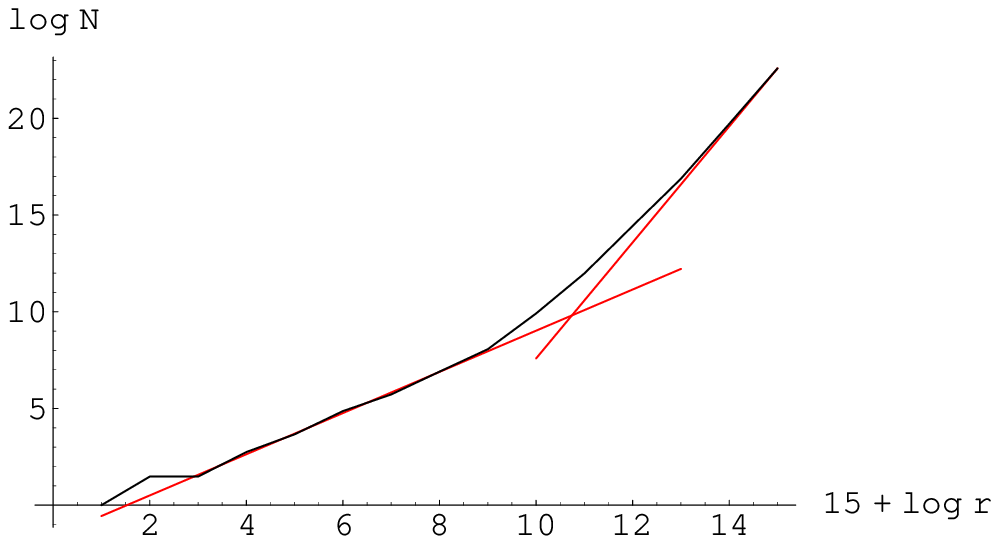}
\includegraphics[width=7.5cm]{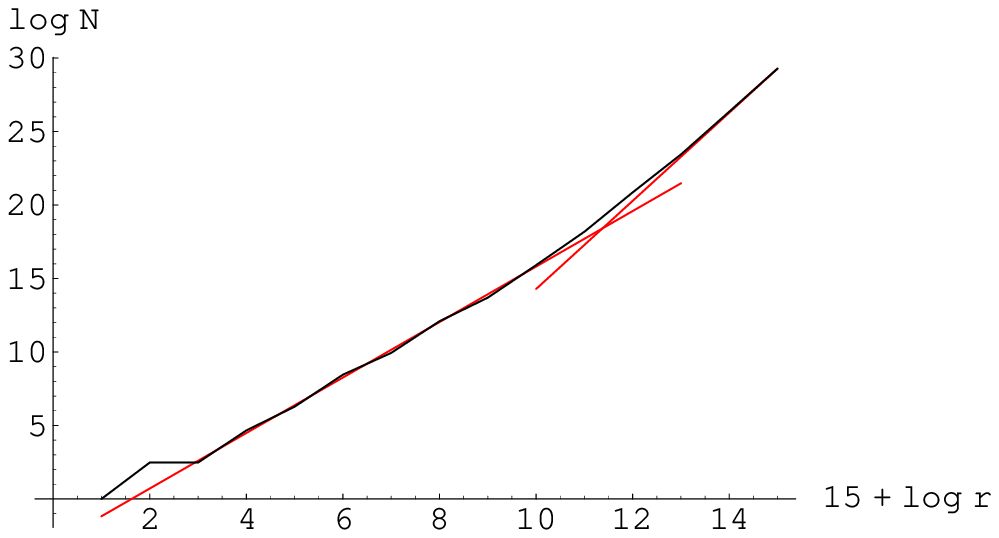}
\end{minipage}
\caption{Selection of two halo populations from the GIF2 $N$-body
simulation: heavy haloes with 750 to 1000 particles and light haloes
with 100 to 150 particles.  (Left) Slice showing heavy haloes in red
and light haloes in blue.  (Right) Number-radius function for the
heavy haloes (above) and the light haloes (below), showing fractal
dimensions $D = 1.1$ and $D = 1.9$, respectively, and a transition to
homogeneity in both (logarithms are to base $\sqrt{2}$ and the total
size is normalized to unity).  }
\label{figs2}
\end{figure}

We have applied this new picture of multifractal models to $N$-body
cosmological simulations by the Virgo Consortium; in particular, we
have taken the $z=0$ positions of the $\L$CDM GIF2 simulation, with
$400^3$ particles in a volume of (110 $h^{-1}$ Mpc)$^3$ \cite{GIF2}.
To measure the strength of singularities, we use a spatial window, for
several window lengths.
To simplify matters, we select two different
singularity strengths, corresponding to two different halo masses (at
definite window length); hence we classify the haloes in heavy and
light haloes, in a sort of bifractal.  If we were to identify haloes
as the places of galaxy formation, this selection would roughly
correspond to two types of galaxies that have been used in connection
with voids in the galaxy distribution, namely, ``wall'' and ``field''
galaxies.  Results of an analysis 
with window length $r = 2^{-8}$
appear in Fig.~\ref{figs2}: On the
left-hand side, there is a slice showing both halo populations,
consisting of 2508 and 25556 members, with 750 to 1000 particles and
with 100 to 150 particles, respectively.  On the right-hand side, we
have log-log plots of the number-radius functions for the respective
spatial distributions of halo centers. 
They have scaling ranges corresponding to quite
different fractal dimensions, namely, $D_1 = 1.1$ and $D_2 = 1.9$.  We
have also carried out the analysis of moments and derived the
multifractal spectrum. It yields entropy dimension $D(1) = 2.5$ and
shows that the values of the fractal dimensions $f(\a_1) = 1.1$ and
$f(\a_2) = 1.9$ in Fig.~\ref{figs2} correspond (approximately) to
$\a_1 = 1.5$ and $\a_2 = 2.0$, and to $q_1 = 2.3$ and $q_2 = 1.5$.
These noninteger values are relatively close to 
the integer $q=2$, associated with the two-point correlation 
function. The role of the various $M_q$ and, in particular, 
of $M_2$ is further discussed below.

We can observe that the light haloes are somewhat clustered, as
corresponds to a fractal of dimension $\simeq 2$. In fact, more
uniform distributions can be found for even lighter haloes, and the
most uniform distribution corresponds to the largest fractal
dimension, namely, $D(0)$. The estimation of $D(0)$ is affected by the
problem of undersampling, especially for small window length.
However, it seems that $D(0)=3$ is already reached for values of the
window length $r = 2^{-7} \simeq 0.01$, that is, well within the
scaling range. Indeed, in the scale range $2^{-7}$--$2^{-4}$ we have a
stable multifractal spectrum. Note that these scales correspond to
physical length $\simeq 1$--$7\; h^{-1}$ Mpc, larger than the size of
virialized haloes.

We remark that $\a(q=0) \simeq 3.3 > 3$, such that the distribution at
the corresponding points is regular with vanishing density. So there
are no haloes at those points, which actually belong to voids.  
The set of regular points with
non-vanishing density corresponds to $\a = 3$ and $f(\a) \simeq 2.9$,
which is nearly homogeneous.

\section{Conclusions}

We have seen, in a model of the large-scale matter distribution
consisting of randomly placed haloes with power-law profile, that the
power-law form of correlation functions as $r \ra 0$ is due to the
singularities of the power-law distribution, rather than to its
regular part. We have also seen that it is only a particular and
degenerate case of a multifractal distribution.  While a suitable
spectrum of singularity strengths can reproduce the scalings observed
in the galaxy distribution \cite{Mur}, underdense regions
are necessarily homogenous and there are no voids. Therefore, we
consider the generalization to a multifractal model consisting of a
set of fractal distributions of power-law haloes.  A degenerate model
with a set of discrete singularity strengths is useful conceptually
and to analyse $N$-body simulations. We have analysed in the GIF2
simulation the distributions of singularities with several
strengths. They show well-defined scaling ranges, with different
scaling dimension, and their transition to homogeneity.  We have
selected two representative populations of heavy or light haloes, with
fractal dimensions $D = 1.1$ and $D = 1.9$. We remark that these two
values span the usual range of fractal dimensions of the galaxy
distribution.

As to the r\^ole of the two-point correlation, it is certainly not
sufficient to fully characterize the scaling properties of the dark
matter distribution and one has to study higher order
correlations. This is characteristic of multifractals, which are
defined by the full set of moments $M_q(r)$ for $-\infty < q <
\infty$, as opposed to pure fractals, defined only by $M_2(r)$ (in
fact, by $M_q \sim M_2^{q-1}$, with any $q$).  In addition, we remark
that the singular nature of multifractals implies that the moments
with $q = 2,3, \ldots$ are not sufficient.  The restriction to these
moments amounts to considering a multifractal degenerate by pieces,
but with an inadequate division.  In fact, most important is $q \simeq
1$, because the entropy dimension $D(1) = \a = f(\a)$ defines the set
of the measure's concentrate.  Larger $q$'s are also relevant but very
little information is actually gathered from moments with $q > 6$ (in
current $N$-body simulations).  Negative-$q$ moments are relevant as
well, but not in connection with haloes but with voids.  At any rate,
if we focus on a uniform set of haloes, namely, on a definite set of
singularities with similar strength, the two-point correlation
function of their centers provides the necessary information for its
scaling properties. Since this is true for every set of singularities,
the corresponding set of two-point correlation functions defines the
scaling properties of the full distribution. Precisely this is the
basis of the picture of the full distribution as a set of fractal
distributions of haloes that we propose.

\section{Acknowledgments}
I am grateful to Liang Gao for kindly supplying me with the GIF2 data
and to Marco Montuori for correspondence regarding their paper
\cite{BMP}. My work is supported by the ``Ram\'on y Cajal'' program
and by grant BFM2002-01014 of the Ministerio de Edu\-caci\'on y
Ciencia.


\begin{thebibliography}{0}

\bibitem{Pee}
\Name{Peebles, P.J.E.}
\REVIEW{A\&A}{32}{1974}{197}

\bibitem{McC-Silk}
\Name{McClelland, J. \and Silk, J.}
\REVIEW{ApJ}{217}{1977}{331}

\bibitem{Sheth-J}
\Name{Sheth, R.K. \& Jain, B.}
\REVIEW{MNRAS}{285}{1997}{231}

\bibitem{Mur}
\Name{Murante, G., Provenzale, A., Spiegel, E.A., Thieberger, R.}
\REVIEW{MNRAS}{291}{1997}{585}

\bibitem{BMP}
\Name{Botaccio, M., Montuori M., Pietronero, L.}
\REVIEW{Europhysics Letters}{66}{2004}{610}

\bibitem{CoorShet} 
\Name{Cooray, A. \& Sheth, R.}
\REVIEW{Phys. Rep.}{372}{2002}{1}

\bibitem{Mo-W}
\Name{Mo, H.J. \& White, S.D.M.}
\REVIEW{MNRAS}{282}{1996}{347}

\bibitem{Sheth-Tor}
\Name{Sheth, R.K. \& Tormen, G.}
\REVIEW{MNRAS}{308}{1999}{119}

\bibitem{Falc} 
\Name{Falconer, K.}
\Book{Fractal geometry}
\Publ{John Wiley \& Sons, Chichester, UK}
\Year{2003}

\bibitem{Halsey}
\Name{Halsey, T.C., Jensen, M.H., Kadanoff, L.P., Procaccia, I. 
\& Shraiman, B.I.}
\REVIEW{Physical Review A}{33}{1986}{1141}

\bibitem{Bal-Schaf}
\Name{Balian, R. \and Schaeffer, R.}
\REVIEW{A\&A}{226}{1989}{373}

\bibitem{Borga}
\Name{Borgani, S.}
\REVIEW{MNRAS}{260}{1993}{537}

\bibitem{adhesion} 
\Name{Gurbatov, S.N., Saichev, A.I. \& Shandarin, S.}
\REVIEW{MNRAS}{236}{1989}{385}

\bibitem{Verga} 
\Name{Vergassola, M., Dubrulle, B., Frisch, U. \& Noullez, A.}
\REVIEW{A\&A}{289}{1994}{325}

\bibitem{Vala}
\Name{Valageas, P.}
\REVIEW{A\&A}{347}{1999}{757}

\bibitem{multif1}
\Name{Dom{\'\i}nguez-Tenreiro, R., \& Mart{\'\i}nez, V.J.}
\REVIEW{ApJ}{339}{1989}{L9}

\bibitem{multif2}
\Name{Mart{\'\i}nez, V.J., Jones, B.J., Dom{\'\i}nguez-Tenreiro, R.
\& van de Weygaert, R.}
\REVIEW{ApJ}{357}{1990}{50}

\bibitem{Colom}
\Name{Colombi, S., Bouchet, F.R. \and Schaeffer, R.}
\REVIEW{A\&A}{263}{1992}{1};

\bibitem{Valda}
\Name{Valdarnini, R., Borgani, S. \and Provenzale, A.}
\REVIEW{ApJ}{394}{1992}{422};

\bibitem{Yepes}
\Name{Yepes, G., Dominguez-Tenreiro, R. \and Couchman, H.P.M.}
\REVIEW{ApJ}{401}{1992}{40};

\bibitem{GIF2} 
\Name{Gao, L., White, S.D.M., Jenkins, A., Stoehr, F. \& Springel, V.}
\REVIEW{MNRAS}{355}{2004}{819}

%\bibitem{Piet}
%\Name{Sylos Labini F., Montuori M. \and Pietronero L.}
%\REVIEW{Phys.\ Rept.}{293}{1998}{61};

\end{thebibliography}
\end{document}